\documentclass[conference]{IEEEtran}

\ifCLASSINFOpdf
\else

\fi

\usepackage{amsmath}

\usepackage{algorithmic}

\usepackage{array}

 \usepackage[caption=false,font=normalsize,labelfont=sf,textfont=sf]{subfig}

\usepackage{biblatex} 
\usepackage{authblk}
\usepackage{amsmath}
\usepackage{enumitem}
\usepackage{hyperref} 
\usepackage{graphicx} 
\begin{document}
%

\title{Interleaved One-Shot SPS Performance under Smart DoS Attacks in C-V2X Networks}

\author{
    Zepei Sun
    \\
    Northwestern University \\
    \texttt{zepeisun2025@u.northwestern.edu}
    \and
    Randall Berry \\
    Northwestern University\\
    \texttt{rberry@northwestern.edu}
}

%
%
%

%
%

\markboth{}%
{Shell \MakeLowercase{\textit{et al.}}: Bare Demo of IEEEtran.cls for IEEE Journals}
%



\maketitle

\begin{abstract}
This paper evaluates the performance of the one-shot Semi-Persistent Scheduling (SPS) mechanism in Cellular Vehicle-to-Everything (C-V2X) networks under Denial-of-Service (DoS) smart attack scenarios. The study focuses on the impact of these attacks on key performance metrics, including Packet Delivery Ratio (PDR), Inter-Packet Gap (IPG), and Age of Information (AoI). Through extensive Monte Carlo simulations, we demonstrate that the one-shot mechanism significantly enhances network resilience by mitigating the adverse effects of smart DoS attacks. The findings reveal that while the one-shot mechanism improves the PDR and reduces the IPG and AoI tail values, its effectiveness diminishes slightly in high-density vehicular environments. Nevertheless, the one-shot mechanism proves to be a robust solution for maintaining the stability and reliability of C-V2X communications under adversarial conditions.
\end{abstract}


%
\IEEEpeerreviewmaketitle

\section{Introduction}

%
%
%
%
\IEEEPARstart{V}{e}hicular communication systems play a crucial role in enhancing the safety, efficiency, and reliability of modern transportation networks. Among the leading technologies in this domain is Cellular Vehicle-to-Everything (C-V2X), which builds upon existing cellular-based LTE features to facilitate the exchange of V2X messages. C-V2X enables direct communication between vehicles (V2V), between vehicles and infrastructure (V2I), vehicles and pedestrians (V2P), and vehicles and the network (V2N) [1]. Initially defined as LTE V2X in 3GPP Release 14, C-V2X operates across several modes: V2V, V2I, V2N, and V2P, enabling the timely exchange of critical information such as safety warnings and traffic updates, thereby reducing accidents and optimizing traffic flow. The reliability of vehicular communication is underscored by its potential to prevent accidents, manage traffic congestion, and enhance the overall driving experience.

C-V2X functions in two primary modes: Mode 3 and Mode 4. Mode 3 relies on network scheduling for resource allocation, where a central entity, typically the cellular network, assigns resources to vehicles. In contrast, Mode 4 allows vehicles to autonomously manage resources, offering greater scalability and flexibility. Further enhancements for advanced New Radio V2X (NR-V2X) were introduced in Release 16 [2]. It is anticipated that support for advanced V2X application use cases will continue to evolve in Release 17 and subsequent releases [3]. However, the decentralized nature of Mode 4 introduces significant vulnerabilities, particularly to interference and malicious attacks. Among these, Denial-of-Service (DoS) attacks are especially concerning, as they can severely disrupt communication by causing packet collisions and reducing the Packet Delivery Ratio (PDR). Trkulja et al. [4] identified several types of DoS attacks, including oblivious, smart, and cooperative attacks. These attacks can significantly degrade network performance, leading to increased packet loss, higher latency, and ultimately, the failure to deliver critical safety messages.

The SAE J3161/1 standard introduces probabilistic one-shot transmissions to reduce the likelihood of consecutive packet collisions in C-V2X, which can sometimes arise due to half-duplex operation and periodic semi-persistent scheduling (SPS) transmissions. Research efforts have been made to assess the potential benefits of one-shot transmission. In studies [5], [6] and [7], Fouda and Berry examined the tail performance of Basic Safety Message (BSM) Inter-Packet Gap (IPG) using one-shot transmission, demonstrating its effectiveness in improving performance. Similarly, Ghomieh et al. [8] concluded that one-shot transmission can enhance IPG tail performance across various traffic density scenarios.

However, despite its potential in improving IPG tail performance, the robustness of the one-shot mechanism under DoS smart attack scenarios remains underexplored. Evaluating its performance in such contexts is crucial for developing resilient vehicular communication systems. Also, while previous research on DoS attacks has primarily focused on basic SPS, advanced mitigation strategies like the one-shot mechanism have not been thoroughly investigated. This paper aims to fill this gap by assessing the effectiveness of the one-shot mechanism in the presence of DoS smart attacks.

The results in [4] show the impact of DoS smart attacks on PDR, but the effects of these attacks on IPG and Age of Information (AoI) in C-V2X systems have yet to be explored. To address this gap, we investigate the impact of DoS smart attacks on IPG and AoI and evaluate the performance of the one-shot mechanism for BSM transmissions in C-V2X Mode 4 networks. Through extensive simulations, we demonstrate that the one-shot mechanism significantly mitigates the adverse effects of smart DoS attacks, thereby enhancing the reliability and resilience of C-V2X communications.

The remainder of this paper is organized as follows: Section II reviews the system model related to DoS attacks in C-V2X networks and the one-shot mechanism. Section III describes the simulation setup and methodologies employed for evaluating the one-shot mechanism under attack scenarios. Finally, Section IV concludes the paper, highlighting the performance improvements achieved through the one-shot mechanism.


\section{System Modeling and C-V2X}

\subsection{Semi-Persistent Scheduling}
In C-V2X transmission mode 4, Vehicle User Equipments (VUEs) use a sensing-based SPS scheme to autonomously allocate radio resources without the need for assistance from the cellular infrastructure. VUEs randomly select the necessary Virtual Resource Blocks (VRBs) for BSM transmission from a candidate list of VRBs. This candidate list is defined within a pre-configured resource pool known as the selection window. The total number of available VRBs for reselection must comprise at least 20$\%$ of all resources within the selection window resource pool.

 When a new BSM is generated by a VUE, the VUE establishes the selection window and selects the same sub-frame within that window for the subsequent consecutive $C_s$ BSM transmissions. The parameter $C_s$, known as the resource reselection counter (or SPS interval), is chosen uniformly at random within the range $[\alpha,\beta]$, where $\alpha$ and $\beta$ are fixed integers satisfying $0<\alpha<\beta$. Once a VUE selects a set of VRBs, it persistently reuses them for the next consecutive several BSM transmissions. Reusing the same VRBs means transmitting in the same sub-frame within each selection window, determined based on the BSM generation time nn and the Packet Delay Budget (PDB) length.The resource reselection counter is decremented by one after each BSM transmission. When Cs reaches zero, the VUE reselects a new set of VRBs with a reselection probability $p_r =1 - p_k$. Here, $p_k$ represents the probability of retaining the current VRBs for the next BSM transmission after Cs reaches zero, with $p_k$ varying between 0 and 0.8 in steps of 0.2. Once new VRBs are selected, or the current VRBs are retained, a new SPS interval begins.

\subsection{One Shot Transmission}
This section outlines the fundamental concept of the one-shot transmission scheme. Typically, one-shot transmissions introduce additional randomness into the resource reselection process to prevent long IPGs (e.g., persistent packet collisions) [5]. Let $C_o$ represent the one-shot resource reselection counter, which is selected uniformly at random within the range $[\rho, \sigma]$, where $\rho$ and $\sigma$ are fixed integers such that $0 < \rho < \sigma$. When one-shot transmissions are employed, the VUE decreases both $C_s$ and $C_o$ by one with each packet transmission.

The implementation of one-shot transmissions is then examined by considering three possible scenarios based on the relationship between $C_s$ and $C_o$.

First, when $C_s$ reaches zero while $C_o > 0$, the VUE again uses $p_r$ to decide whether to reselect a new set of VRBs. If a new set of VRBs is selected, the VUE resets both counters and starts the process anew. If not, the VUE only resets the SPS counter $C_s$ and decreases $C_o$ by one.

Second, when $C_o$ reaches zero, a new set of resources is reselected and used only for the current transmission opportunity. The one-shot VRBs are selected using the same sensing-based reselection process discussed earlier. The VUE then resets $C_o$ and uses the regular SPS-granted VRBs for the next BSM transmission opportunity.

Finally, when both $C_s$ and $C_o$ reach zero simultaneously, the VUE resets both counters and uses $p_r$ to decide whether to reselect a new set of VRBs. If a new set of VRBs is selected, the VUE continues to use it for BSM transmission until either $C_s$, $C_o$, or both expire, at which point the process repeats.

In this manner, the old SPS-granted VRBs are utilized for BSM transmission in the next opportunity unless reselected. If reselected, the new VRBs are used until the counters once again reach zero, repeating the cycle as needed.

\subsection{DoS Smart Attack}
In this section, we examine the DoS smart attack as outlined in [4]. This attack strategy targets resource blocks exclusively utilized by a single vehicle, indicating a successful transmission. In our model, the attackers continuously monitor resource usage over the preceding 1000 ms, compiling a list of resource blocks employed by these single vehicles. From this compiled list, the attackers randomly select blocks to target. However, due to the absence of coordination among the attackers, they may inadvertently target the same target vehicle within a single attack period.

In our paper, the DoS smart attack is modeled as follows:

For each subchannel \( j \), we define a binary vector \( \mathbf{u}_j \) of length \( T \), where:
\[
\mathbf{u}_j = [u_j(1), u_j(2), \dots, u_j(T)]
\]
In this vector, \( u_j(t) = 1 \) indicates the resource block \( j \) was utilized during transmission time \( t \), while \( u_j(t) = 0 \) denotes that resource block was not used at that time.

The total usage of resource block \( j \) over the last 1000 milliseconds is computed as:
\[
U_j = \sum_{t=1}^{T} u_j(t)
\]
Here, \( U_j \) represents the total number of transmission intervals during which resource block \( j \) was active within the last 1000 milliseconds.

The attacker then constructs a target set \( S \) consisting of resources that have been utilized at least once within the last 1000 milliseconds:
\[
S = \{ j \mid U_j > 0, \text{ for } j = 1,2,\ldots,M \}
\]
This set \( S \) includes all resources that have recently been in use by any vehicle.

From the target set \( S \), the attacker randomly selects a resource \( j_{\text{attack}} \) according to a uniform distribution:
\[
j_{\text{attack}} \sim \text{Uniform}(S)
\]

After selecting a resource, the attackers adhere to the SPS scheme, which requires them to maintain the selected resource for a predetermined interval before reselecting a new one. However, for DoS Smart attackers, the key distinction is that upon the conclusion of the attack, they will always reselect a resource rather than doing so with a probability $\rho$.





\section{Performance Evaluation}
In this section, we present the results of Monte Carlo simulations to assess the potential performance enhancements provided by the one-shot mechanism in an attack environment. Specifically, our study seeks to answer the following research questions:

\begin{enumerate}
    \item \textbf{Packet Delivery Ratio}:
    Fouda et al. [3] demonstrate that the one-shot mechanism can negatively impact the PDR. In an attack environment, We aim to determine whether the one-shot mechanism can indeed enhance PDR in an attack scenario.
        
    \item \textbf{Inter-Packet Gap}:
    We evaluate the potential improvements in IPG provided by the one-shot mechanism under attack conditions, and quantify these enhancements.
    
    \item \textbf{Age of Information}:
   We investigate the impact of the one-shot mechanism on AoI within an attack context and establish the extent of improvement.
    
    \item \textbf{Influence of Attack Interval on One-Shot Performance}:
    We analyze how variations in the attack interval influence the efficacy of the one-shot mechanism.
    \
\end{enumerate}

\textbf{\textit{Simulation Setup}}

To simulate the attack scenarios, we utilize a Monte Carlo simulation model for C-V2X networks, as introduced in [9], using Python. The simulations are conducted under the following assumptions and conditions, unless otherwise specified in the captions of the figures:

\begin{itemize}
    \item The vehicular network is fully connected, ensuring that all vehicles are within communication range of each other.
    \item There is no signal fading and packet losses are only due to collisions.
    \item Simulation time is 2000000 seconds.
    \item Transmission period, $T_{\text{tr}}$, is set to 100 ms.
    \item The Semi-Persistent Scheduling interval is (5,15)
    \item The one-shot interval is (2,6) or (5,15)
    \item There are 100 resource blocks ($N_r = 100$).
    \item The probability that a target vehicle changes its resource block is $p = 0.2$.
\end{itemize}

\subsection{Packet Delivery Ratio}
In this section, we analyze the PDR performance when one-shot reselection is used for BSM transmission. PDR is a crucial performance metric for evaluating transmission reliability in C-V2X networks. Specifically, we calculate PDR under varying vehicle densities. The PDR is determined by the ratio \( R/T \), where \( R \) denotes the number of successfully received packets, and \( T \) represents the total number of transmission attempts across all transmitter-receiver pairs. Fig. 1 illustrates the PDR as a function of the number of target vehicles, with the number of attackers fixed at five. First, we analyze the environment with attackers. As anticipated, The PDR remains highest in scenarios without attackers and the scenario with attackers and no one-shot mechanism exhibits the lowest PDR. When the number of target vehicles is five, the addition of attackers decreases the PDR by nearly 50$\%$. However, as the number of vehicles increases, the PDR curves for scenarios with and without attackers converge indicating the impact of attackers diminishes.           

The implementation of the one-shot (2,6) configuration significantly improves the PDR by 59.62$\%$ when there are five target vehicles. Similarly, the one-shot (5,15) configuration enhances the PDR by 39.22$\%$ under the same conditions. However, As the number of vehicles increases, the benefit of the one-shot mechanism diminishes. In high-density environments, the one-shot (2,6) configuration slightly decreases the PDR by 3.49$\%$, whereas the one-shot (5,15) configuration continues to improve the PDR by 3.34$\%$. Comparing the two one-shot configurations, it is evident that the (2,6) configuration offers better PDR performance in low vehicle density scenarios. However, as vehicle density increases, the performance improvement of the (2,6) configuration declines, eventually becoming inferior to the (5,15) configuration.

Moreover, in the environments without attackers, the (2,6) configuration consistently results in a degradation of PDR across all scenarios. In contrast, the (5,15) configuration begins to enhance the PDR in high-density environments. This indicates that while the (2,6) configuration may provide significant improvements in specific scenarios, the (5,15) configuration delivers more consistent and average improvements in PDR across a broader range of vehicle densities.

\begin{figure}[!h]
		\centering
		\includegraphics[width=0.35\textheight]{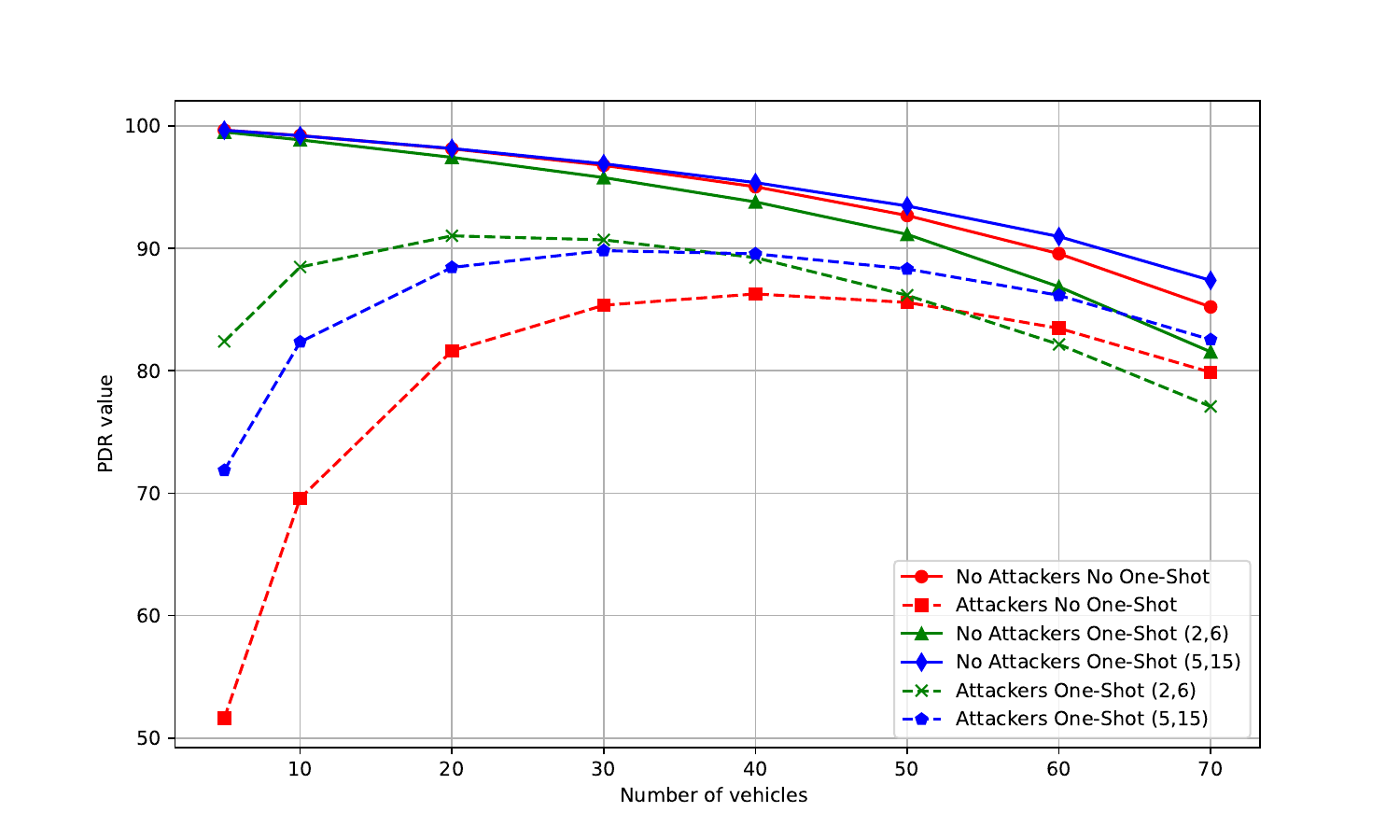} \\
		\label{ann_images}
      \caption{Packet Delivery Ratio}

	\end{figure}

\subsection{Inter-packet Gap}

This section presents an analysis of the IPG in vehicular networks under attack, focusing on the effectiveness of a one-shot mechanism in mitigating the impact of such attacks. For each pair of vehicles, the IPG is measured as the time interval between successive successfully received BSMs. The IPG Complementary Cumulative Distribution Function (CCDF) $F(i)$ represents the proportion of these intervals that exceed $i$ milliseconds. In our paper, the primary focus is on the IPG value at the $10^{-5}$ probability point. The analysis is based on two configurations of the one-shot mechanism: (2,6) and (5,15), and examines their performance across varying vehicle densities.

\subsubsection{$10^{-5}$ IPG tail}
We begin by analyzing the IPG without the implementation of the one-shot mechanism. As illustrated in fig. 2, when the number of target vehicles is 5, the interference caused by the attacker reaches its peak, leading to a significant increase in the IPG tail, nearly doubling compared to the scenario without attackers. However, as vehicle density increases, the effect of the attackers diminishes, indicating that higher vehicle density reduces the overall impact of the interference.

The introduction of the one-shot mechanism results in a marked improvement in IPG performance under attack conditions. The IPG tail is significantly reduced, and the IPG performance remains stable even as the number of vehicles increases. This stability is evident in the comparison of curves, where the one-shot mechanism demonstrates superior resilience against attacks.

For the scenario with 5 target vehicles, the (2,6) configuration improves the IPG tail by approximately 90.37$\%$, while the (5,15) configuration achieves an improvement of around 80.42$\%$. Notably, as the vehicle density increases, the improvements become more pronounced. For instance, when the number of vehicles reaches 70, the IPG tail improvement is nearly 14,000 ms for the (2,6) configuration and approximately 13,000 ms for the (5,15) configuration.

A comparative analysis reveals that the one-shot mechanism provides effective defense against attacker interference. In the case of 5 target vehicles, the presence of attackers without the one-shot mechanism degrades the IPG by nearly double, adding approximately 6,000 ms. However, with the one-shot mechanism, it remains below 1,200 ms, demonstrating significant mitigation of the attack's impact.

Furthermore, the (2,6) configuration consistently outperforms the (5,15) configuration across all scenarios. The curve for the (2,6) configuration under attack is consistently lower than that of the (5,15) configuration, even outperforming the (5,15) configuration in the absence of attackers. This indicates that the (2,6) configuration offers superior IPG tail improvement and is more effective in mitigating the impact of attacks.

Without attackers, the IPG tail increases from approximately 6,000 ms to nearly 16,000 ms as the vehicle density grows from 5 to 70 target vehicles. However, after implementing the one-shot mechanism, the [2,6] configuration keeps the IPG tail below 2500 ms even with 70 target vehicles, while the [5,15] configuration maintains the IPG tail under 3,500 ms. This substantial improvement underscores the effectiveness of the one-shot mechanism in significantly enhancing IPG performance, particularly in high-density vehicular scenarios.

The results demonstrate that the one-shot mechanism, particularly the (2,6) configuration, significantly enhances IPG performance under attack, with its effectiveness increasing as vehicle density rises. The one-shot mechanism not only mitigates the impact of attacker interference but also stabilizes IPG performance across various vehicle densities.

\subsubsection{probability of a 100 ms}
Another metric we investigated is the probability of achieving a 100 ms IPG, which corresponds to the transmission time for each BSM in our model. This metric is defined as the proportion of IPG equal to 100 ms relative to the total number of observed IPG values. The fig. 3 indicates that, in the absence of attackers, the probability of a 100 ms IPG remains consistently high, exceeding 98$\%$. However, when attackers are introduced and the number of target vehicles is 5, this probability decreases from 99.99$\%$ to approximately 94.3$\%$, representing a reduction of 5.3$\%$. Consistent with our earlier discussion, we could observed that the impact of attackers diminishes as the number of vehicles increases.

The introduction of the one-shot mechanism slightly decreases the probability of achieving a 100 ms IPG, regardless of whether the (2,6) or (5,15) configuration is used. For example, with 5 target vehicles, the (2,6) configuration reduces this probability by 0.48$\%$ without attackers and 0.25$\%$ with attackers. Similarly, the (5,15) configuration results in reductions of 1.16$\%$ in the no-attack scenario and 0.1$\%$ in the attack scenario.

As vehicle density increases, the data show that the degradation associated with the one-shot mechanism becomes more pronounced. In scenarios with 70 target vehicles, the (2,6) configuration reduces the probability of a 100 ms IPG by 19$\%$ in the absence of attackers and by 20$\%$ in the presence of attackers. These findings suggest that, in terms of the probability of achieving a 100 ms IPG, the (2,6) configuration performs less effectively compared to the (5,15) configuration, with this performance gap widening as vehicle density increases.

In summary, the (2,6) configuration outperforms the (5,15) configuration in terms of IPG tail reduction but results in a lower probability of achieving a 100 ms IPG. The (2,6) configuration effectively improves the IPG tail but at the cost of reducing the likelihood of packets being received within a single transmission period. On the other hand, the (5,15) configuration provides more balanced performance, offering slightly less improvement in the IPG tail but significantly better results in maintaining the probability of a 100 ms IPG.
\begin{figure}[!h]
		\centering
		\includegraphics[width=0.35\textheight]{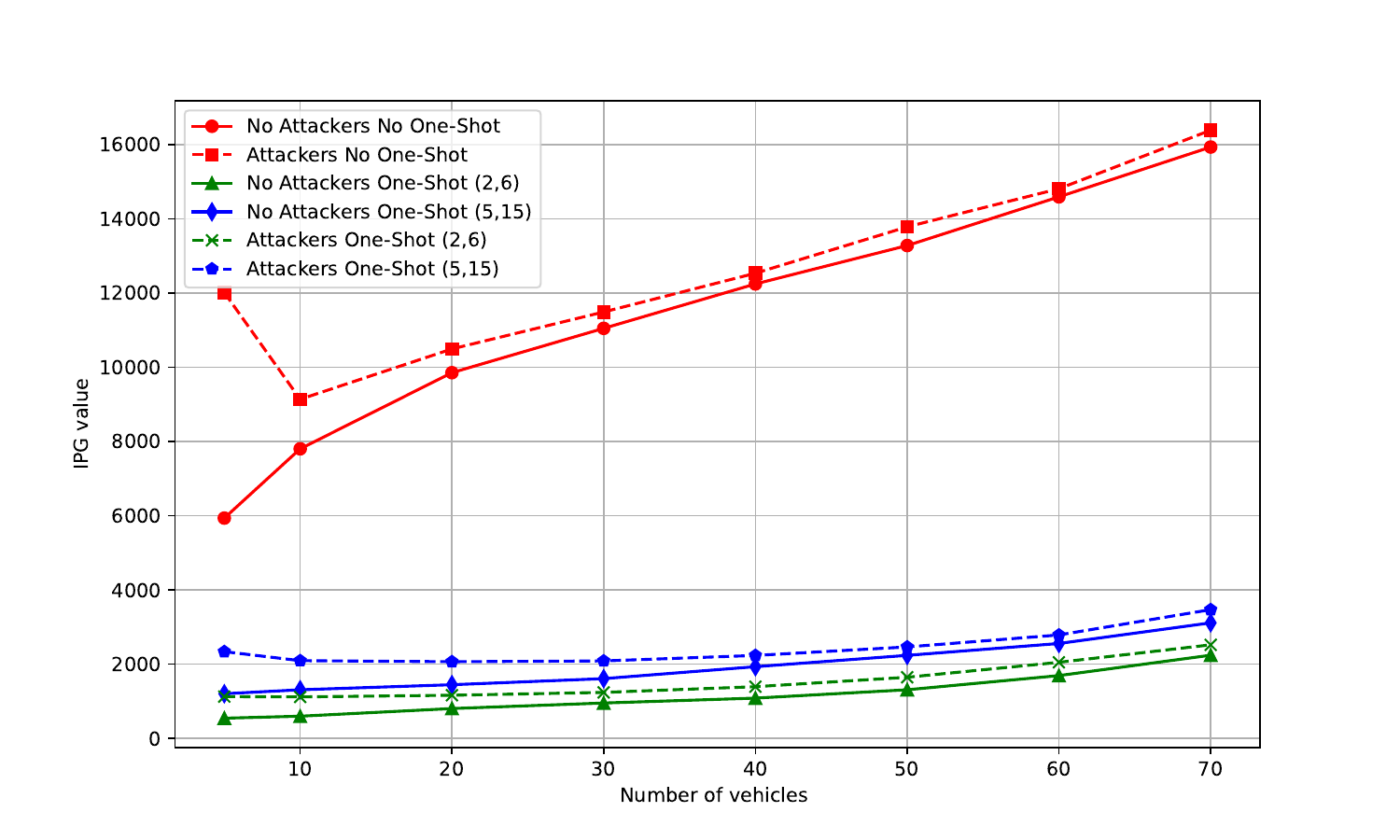} \\
		\label{ann_images}
      \caption{$10^{-5}$ IPG Tail}

	\end{figure}

 \begin{figure}[!h]
		\centering
		\includegraphics[width=0.35\textheight]{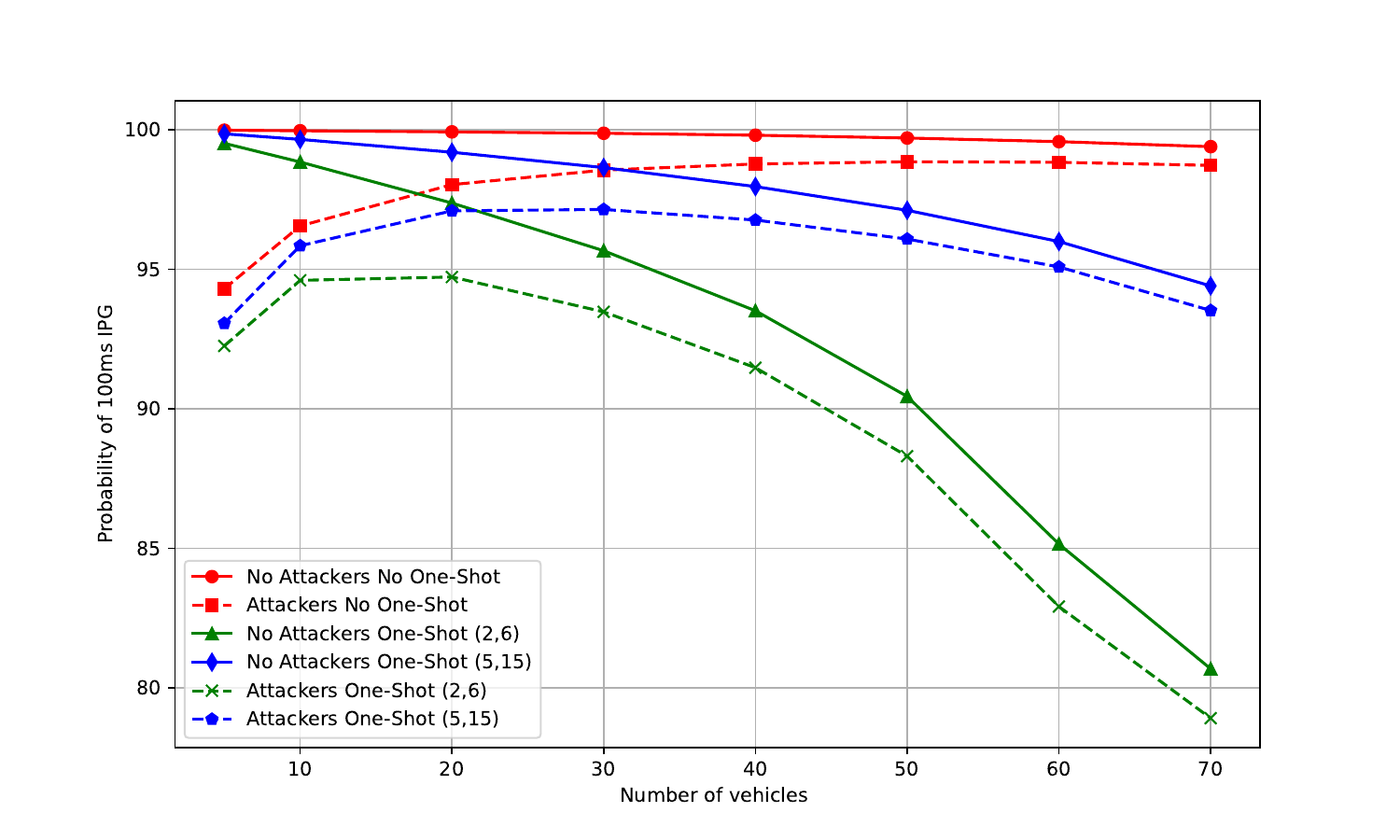} \\
		\label{ann_images}
      \caption{Probability of 100ms IPG}

	\end{figure}
\subsection{Age of Information}

Another important metric in V2X networks related to IPG is Information Age. This metric periodically measures the time elapsed since the generation of the most recently received BSM at the destination VUE. The IA of a BSM received at VUE B from VUE A at time $t_c$ is defined as follows:
$$ I_{B,A} = t_c - t_{s,B}$$

In this context,$t_c$ and $t_s$ denote the current time and the time of the last successfully received BSM at VUE B, respectively. In the proposed model, AoI data is collected based on the difference between the current time and the time of the most recently received BSMs. Specifically, the AoI (CCDF) F(i) represents the proportion of instances where the AoI exceeds i milliseconds.The analysis of AoI tail values under attack scenarios offers critical insights into the impact of adversarial conditions and the effectiveness of the one-shot mechanism. 

\subsubsection{$10^{-5}$ AoI tail}
As observed in Fig. 4, the introduction of attackers does not significantly degrade the AoI tail. For instance, in the scenario with 5 target vehicles, the presence of attackers results in a 16$\%$ degradation of the AoI tail. However, as vehicle density increases, the adverse impact of the attackers diminishes. Although the AoI in an attack scenario remains marginally higher than in a non-attack scenario, the degradation is generally negligible.

The application of the one-shot mechanism leads to significant improvements in AoI tail values. Specifically, in attack environment, for 5 target vehicles, the AoI improves by approximately 92.18$\%$ with the (2,6) configuration and by 83.45$\%$ with the (5,15) configuration. For scenarios involving 40 target vehicles, the AoI improvement is 91.94$\%$ for the (2,6) configuration and 86.96$\%$ for the (5,15) configuration. Similarly, for 70 target vehicles, the AoI improves by 89.81$\%$ with the (2,6) configuration and 84.75$\%$ with the (5,15) configuration. These results clearly demonstrate that the one-shot mechanism significantly enhances AoI performance across both low and high-density environments. However, it is also evident that the degree of improvement decreases as vehicle density increases. In the non-attack environment, similar improvements are observed with the use of the one-shot mechanism. Specifically, the (2,6) configuration yields improvements of 95.71$\%$, 93.48$\%$, and 88.32$\%$ for 5, 40, and 70 target vehicles, respectively. Similarly, the (5,15) configuration results in improvements of 90.58$\%$, 88.05$\%$, and 82.98$\%$ for the same target vehicle scenarios.

Notably, under attack conditions, the (2,6) configuration with the one-shot mechanism outperforms the non-attack scenario utilizing the (5,15) configuration. This finding underscores the superior performance of the (2,6) configuration relative to the (5,15) configuration in terms of AoI tail, highlighting its exceptional effectiveness in mitigating the impact of attacks on AoI.

\begin{figure}[!h]
		\centering
		\includegraphics[width=0.35\textheight]{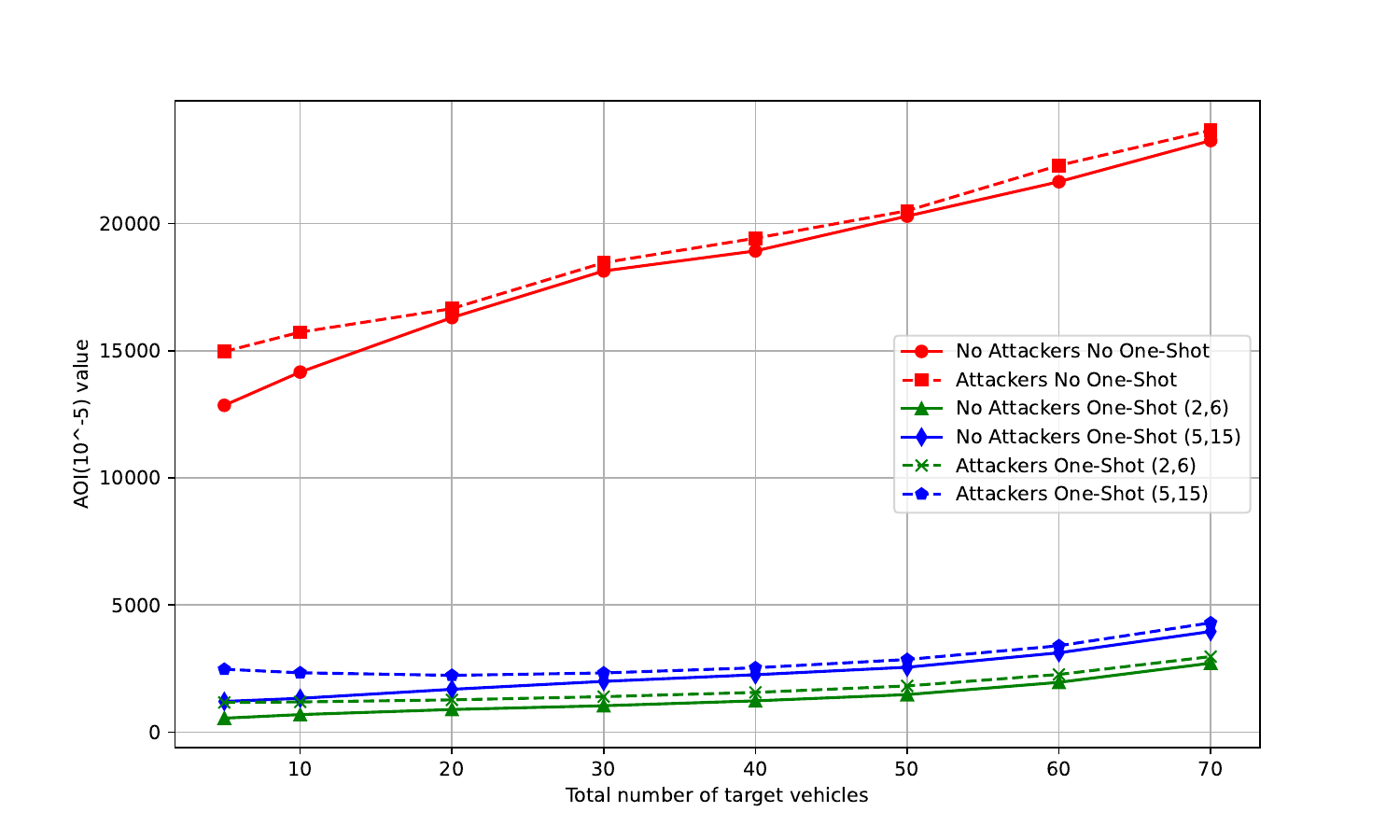} \\
		\label{ann_images}
      \caption{$10^{-5}$ AoI Tail}

	\end{figure}

\subsubsection{probability of a 0 ms AoI}
It is important to note that 0 ms AoI represents the probability of receiving fresh information. In our model, whenever a VUE receives a new packet, the AoI for that VUE is reset to 0 ms. Consequently, the number of instances with 0 ms AoI is directly equivalent to the number of received packets, making the probability of 0 ms AoI identical to the PDR. Therefore, no further analysis of 0 ms AoI is necessary.

\subsection{Impact Analysis of Attack Interval}
In this section, we investigate the effect of the attack interval by analyzing two specific scenarios. The first scenario involves 5 target vehicles and 5 attackers, while the second scenario involves 30 target vehicles and 30 attackers. These scenarios were chosen for the following reasons:
\begin{itemize}
    \item Setting the number of target vehicles equal to the number of attackers allows for a focused examination of the impact of the attack interval.
    \item The first scenario represents a low-density setting, facilitating an analysis of the interval's effect under less congested conditions. Conversely, the second scenario represents a high-density environment, enabling a thorough evaluation of the interval's impact in a more crowded context.
\end{itemize}
\subsubsection{Impact of Attacker Interval on PDR}

The Figs. 5 and 6 illustrate that variations in the attacker interval do indeed influence the PDR, though the impact is not substantial. The most significant attack effects are observed when the interval is either very short or mid-range. This suggests that an attacker achieves more effective disruption when it can rapidly reselect resource blocks to target or when its selection interval aligns with the average SPS behavior of the target vehicles, it can achieve more effective disruption. Conversely, if attackers remain on a selected resource for an extended period, their ability to degrade the PDR diminishes. Additionally, fig. 5 and 6 indicate that the introduction of the one-shot mechanism reduces the influence of the attacker interval on PDR, regardless of whether the scenario involves low or high vehicle density. We can determine this by calculating the difference between the maximum and minimum points. In the 5-target scenario, the difference for the (2,6) and (5,15) configurations is approximately 1.7$\%$ and 3.5$\%$, respectively, whereas for the no one-shot scenario, the difference is 6.6$\%$. Thus, it can be concluded that the one-shot mechanism effectively mitigates the impact of attacks on PDR across various attacker interval settings.
\begin{figure}[!h]
		\centering
		\includegraphics[width=0.35\textheight]{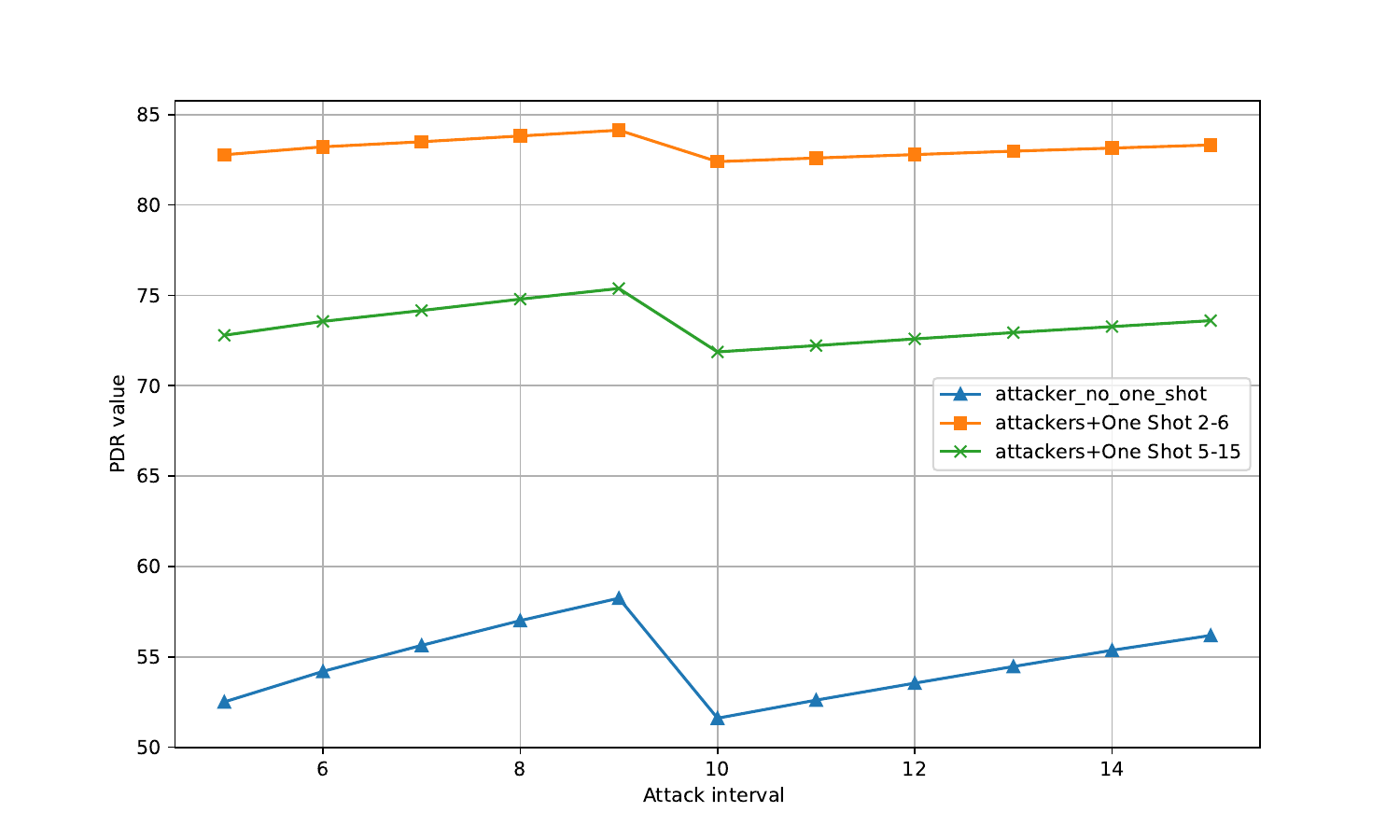} \\
		\label{ann_images}
      \caption{Impact of Attacker Interval on PDR, 5 targets }

	\end{figure}
 \begin{figure}[!h]
		\centering
		\includegraphics[width=0.35\textheight]{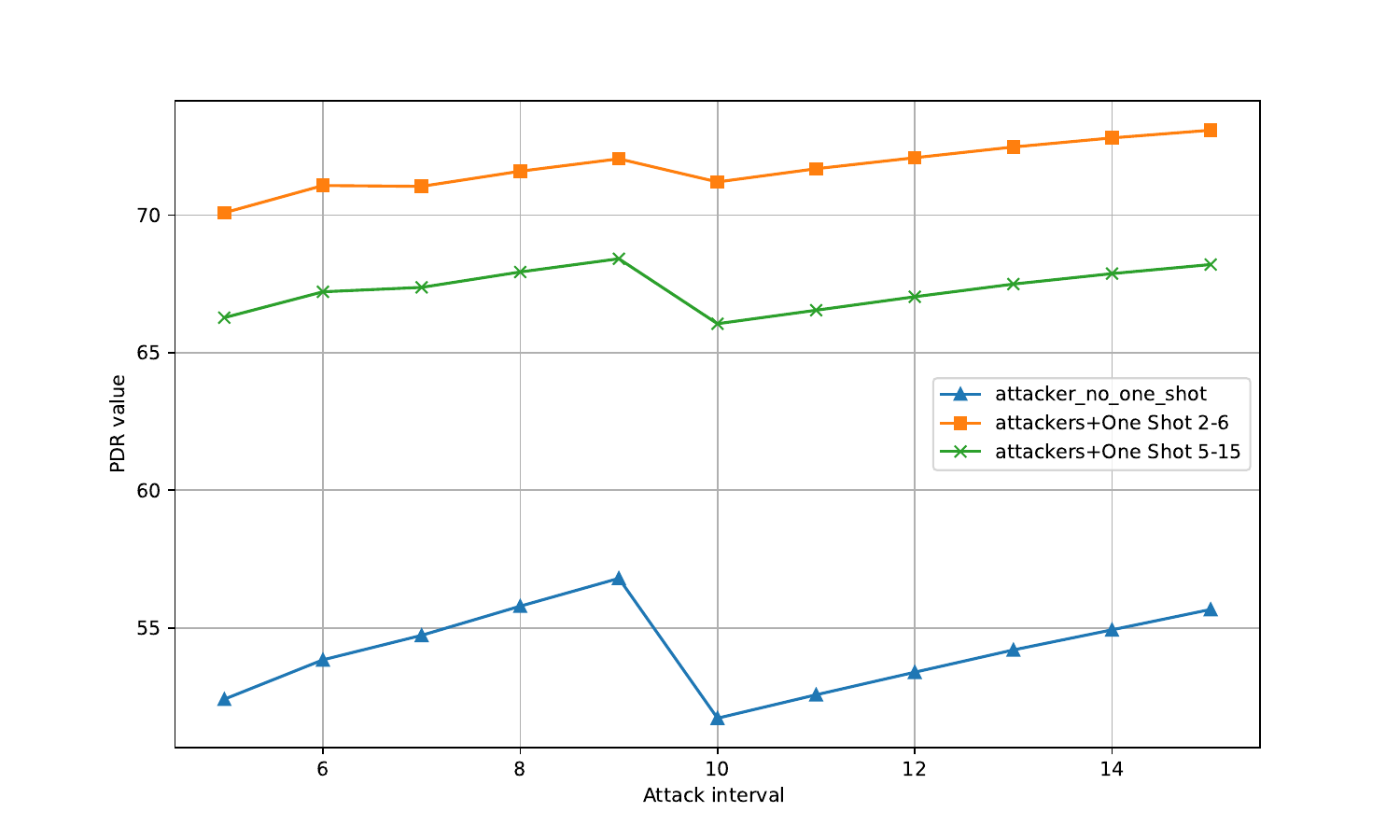} \\
		\label{ann_images}
      \caption{Impact of Attacker Interval on PDR, 30 targets}

	\end{figure}

\subsubsection{Impact of Attacker Interval on $10^{-5}$ IPG tail}

\begin{figure}[!h]
		\centering
		\includegraphics[width=0.35\textheight]{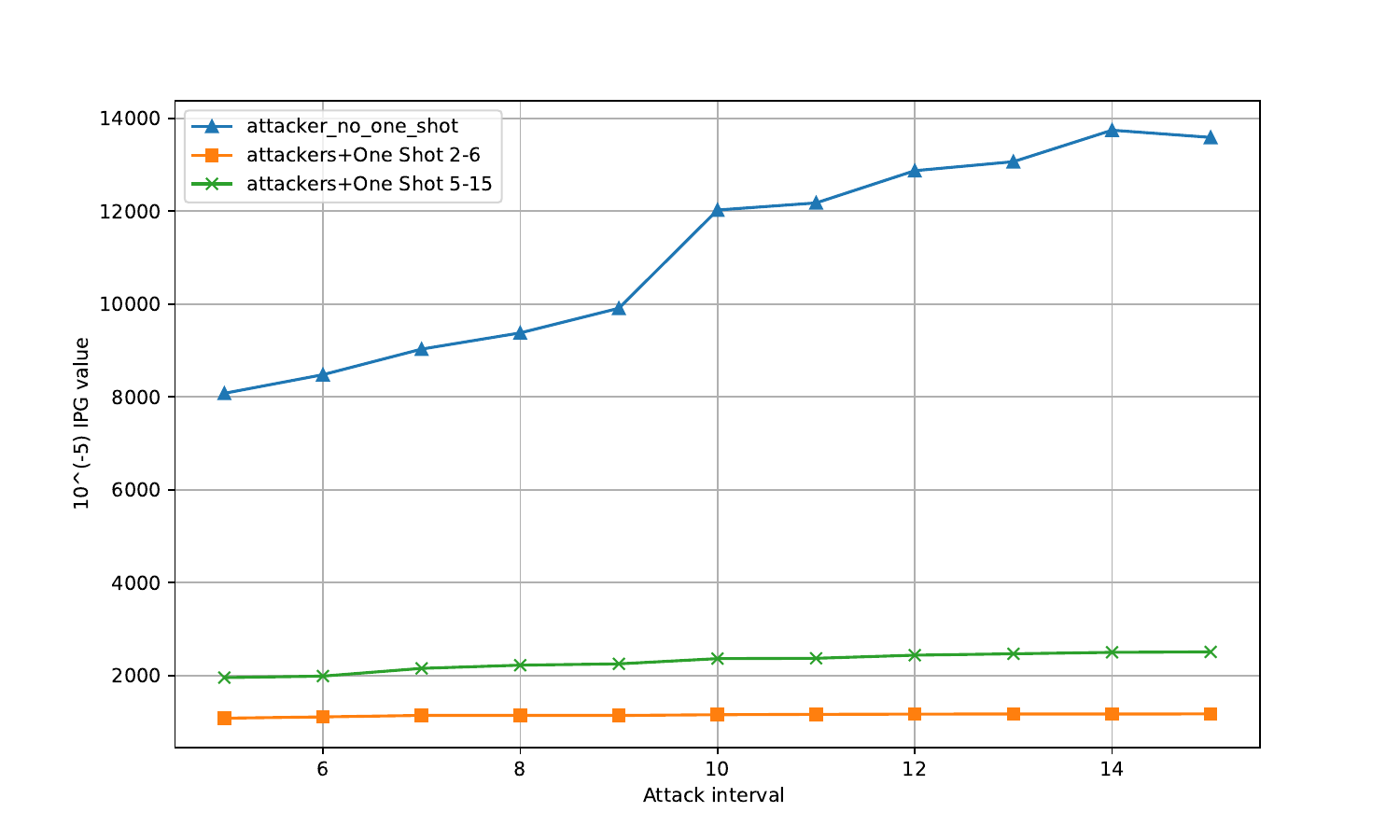} \\
		\label{ann_images}
      \caption{Impact of Attacker Interval on $10^{-5}$ IPG tail, 5 targets}

	\end{figure}

 \begin{figure}[!h]
		\centering
		\includegraphics[width=0.35\textheight]{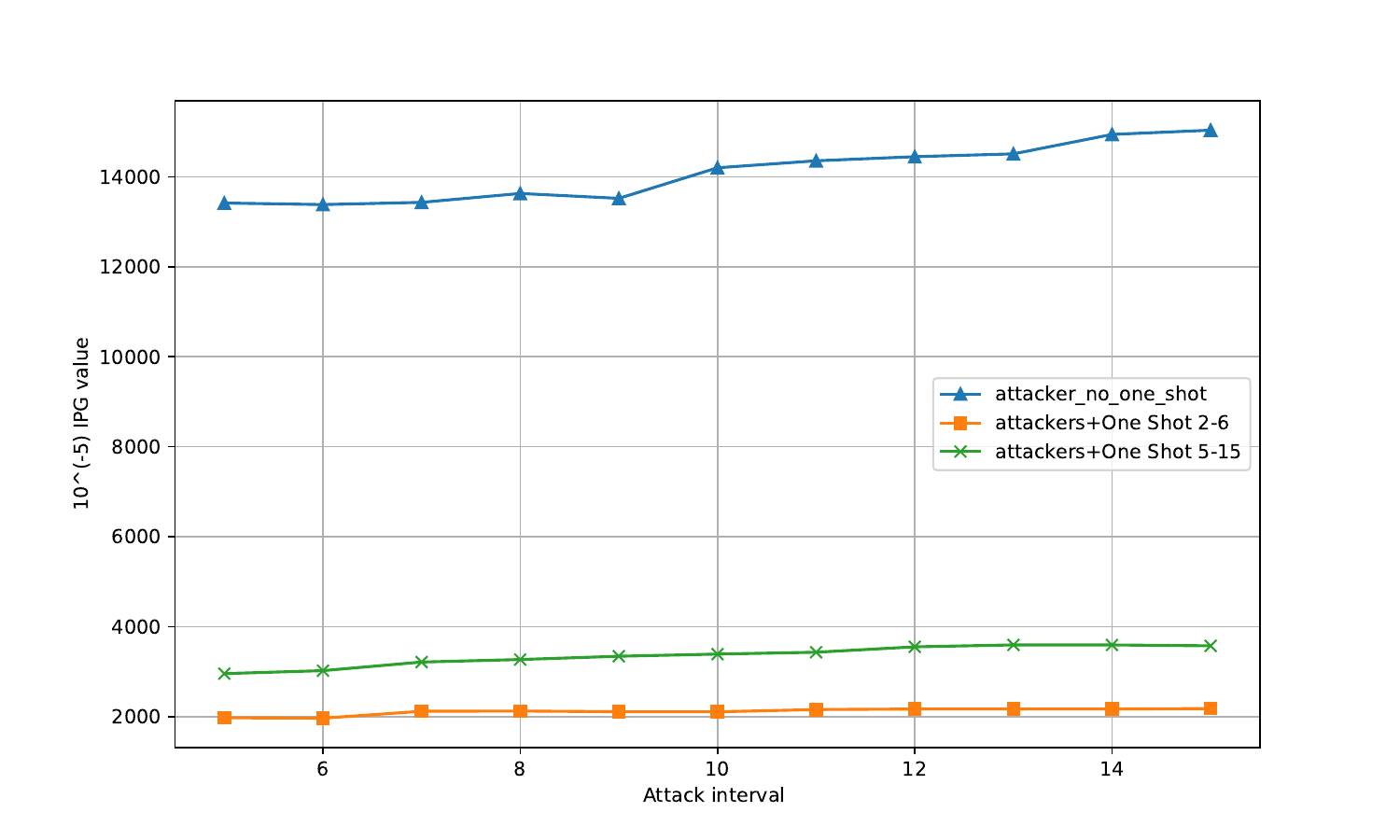} \\
		\label{ann_images}
      \caption{Impact of Attacker Interval on $10^{-5}$ IPG tail, 30 targets}

	\end{figure}

In this analysis, we examine the impact of attack intervals on the $10^{-5}$ IPG tail. Fig. 7 indicates that in low-density scenarios, the attack interval significantly influences the $10^{-5}$ IPG tail, with the tail length increasing from approximately 8000 ms to 14000 ms as the attack interval lengthens.  this trend does not persist in high-density scenarios, as illustrated in Fig. 8, where the  $10^{-5}$ IPG tail remains steady at around 14000 ms, even when the attack interval reaches the maximum of the SPS interval. Additionally, the introduction of the one-shot mechanism proves effective in mitigating the impact of attack intervals. In both high- and low-density scenarios, and across both the (2,6) and (5,15) one-shot configurations, the $10^{-5}$ IPG tail remains stable, showing minimal variation despite changes in the attack interval.

\subsubsection{Impact of Attacker Interval on Probability of 100 ms IPG}
In this part, we examine the impact of the attack interval on the probability of achieving a 100 ms IPG. Figs. 9 and 10 indicate that, in both low-density and high-density scenarios, varying the attack interval significantly influences the probability of a 100 ms IPG. Specifically, when the attack interval is short, the probability of achieving a 100 ms IPG is relatively low; however, as the attack interval increases, this probability improves rapidly. Additionally, while the introduction of the one-shot mechanism slightly reduces the probability of a 100 ms IPG, it also contributes to stabilizing the probability across different attack intervals.

\begin{figure}[!h]
		\centering
		\includegraphics[width=0.35\textheight]{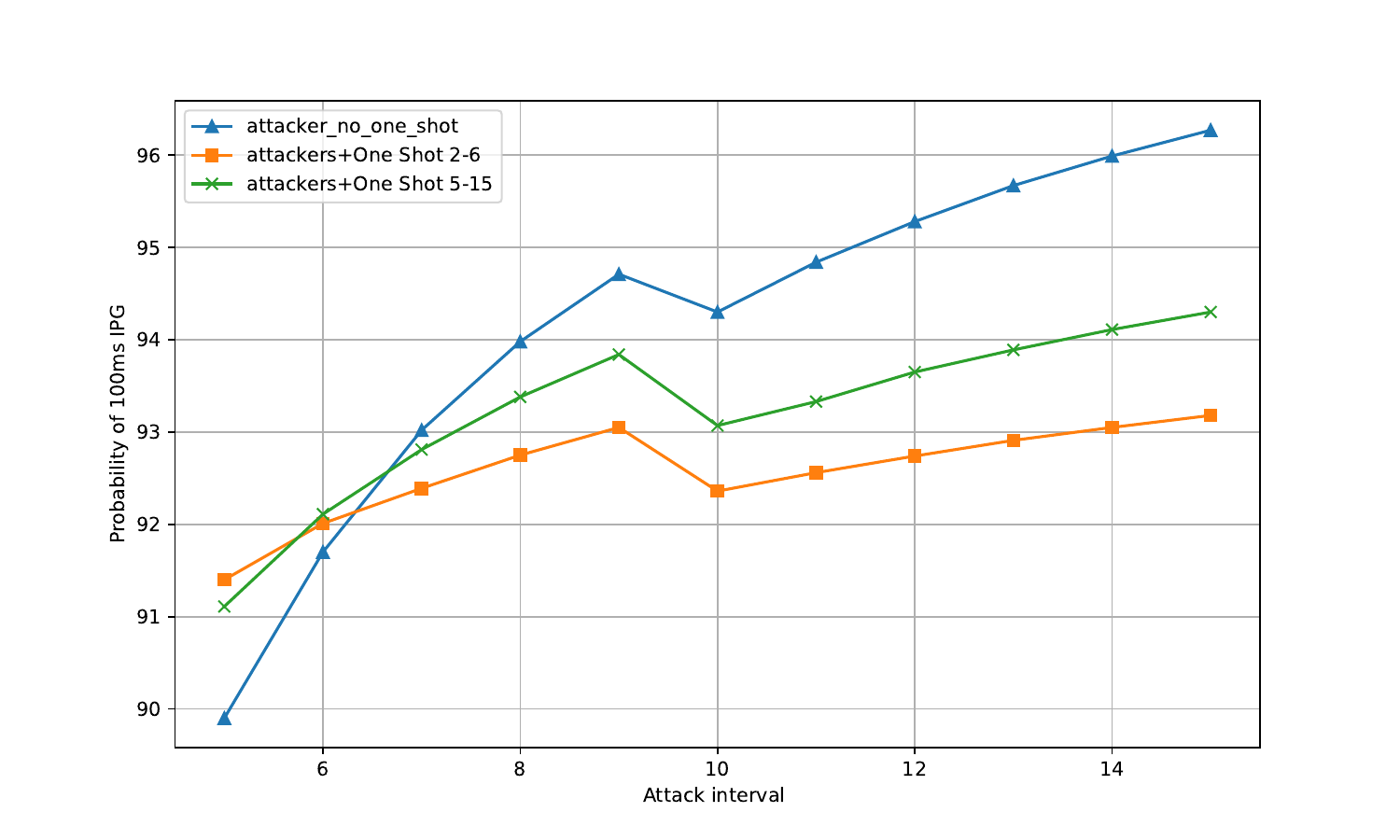} \\
		\label{ann_images}
      \caption{Impact of Attacker Interval on Probability of 100ms IPG, 5 targets}

	\end{figure}

 \begin{figure}[!h]
		\centering
		\includegraphics[width=0.35\textheight]{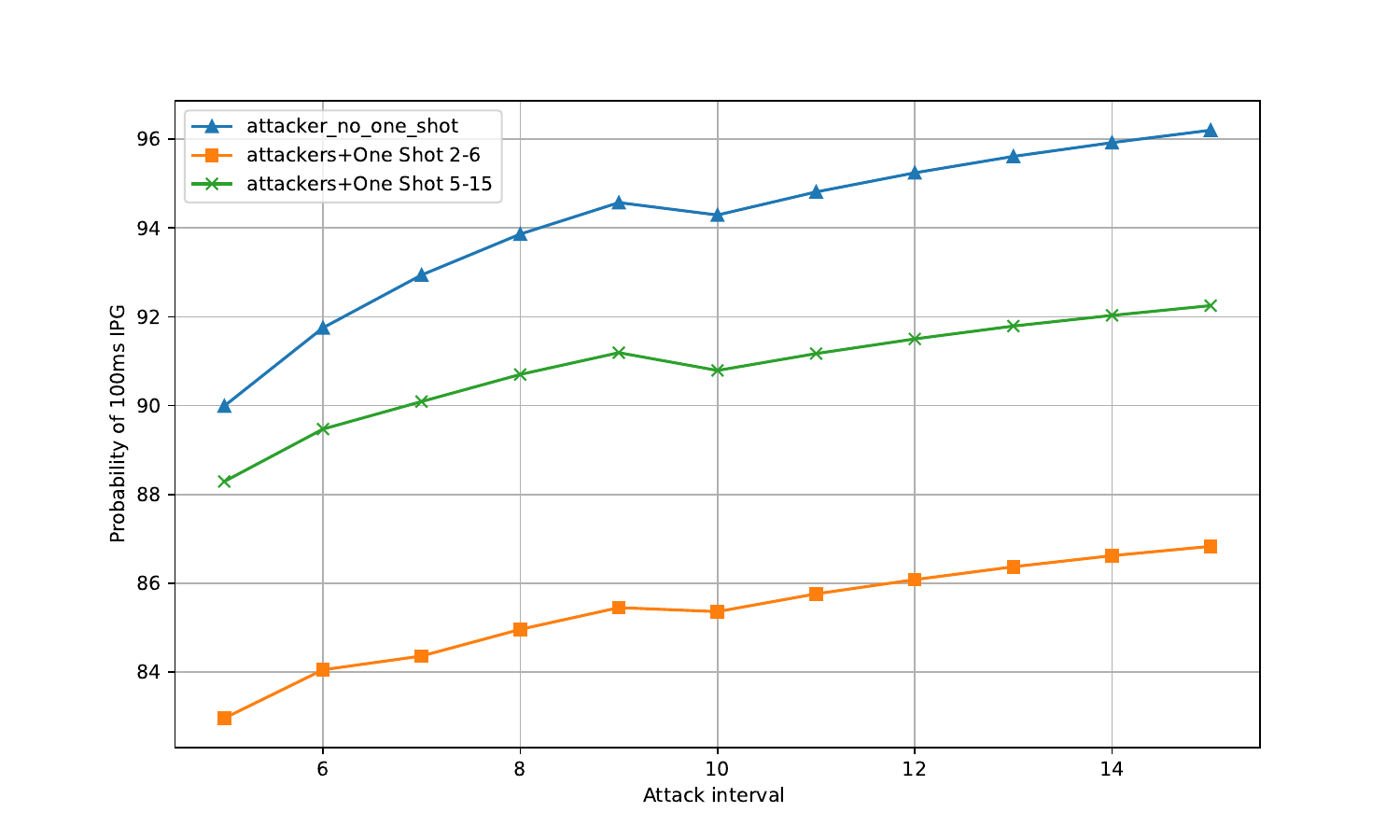} \\
		\label{ann_images}
      \caption{Impact of Attacker Interval on Probability of 100ms IPG, 30 targets}

	\end{figure}
\subsubsection{Impact of Attacker Interval on $10^{-5}$ AoI tail}

In this section, we analyze the impact of the attack interval on the probability of the $10^{-5}$ AoI tail. Fig. 10 and 11 reveal a pattern similar to that observed for the IPG tail. In the scenario with 5 target vehicles, the $10^{-5}$ AoI tail exhibits significant fluctuations as the attack interval varies, with the peak occurring at the largest attack interval. However, it is evident that after implementing the one-shot mechanism, the $10^{-5}$ AoI tail remains nearly constant across different attack intervals for both the (2,6) and (5,15) configurations. These findings indicate that the one-shot mechanism plays a critical role in stabilizing the $10^{-5}$ AoI tail under varying attack behaviors in low vehicle density scenarios. In high vehicle density scenarios, while the $10^{-5}$  AoI tail does not fluctuate as markedly as in the low-density scenario, the application of the one-shot mechanism still effectively reduces volatility.
\begin{figure}[!h]
		\centering
		\includegraphics[width=0.35\textheight]{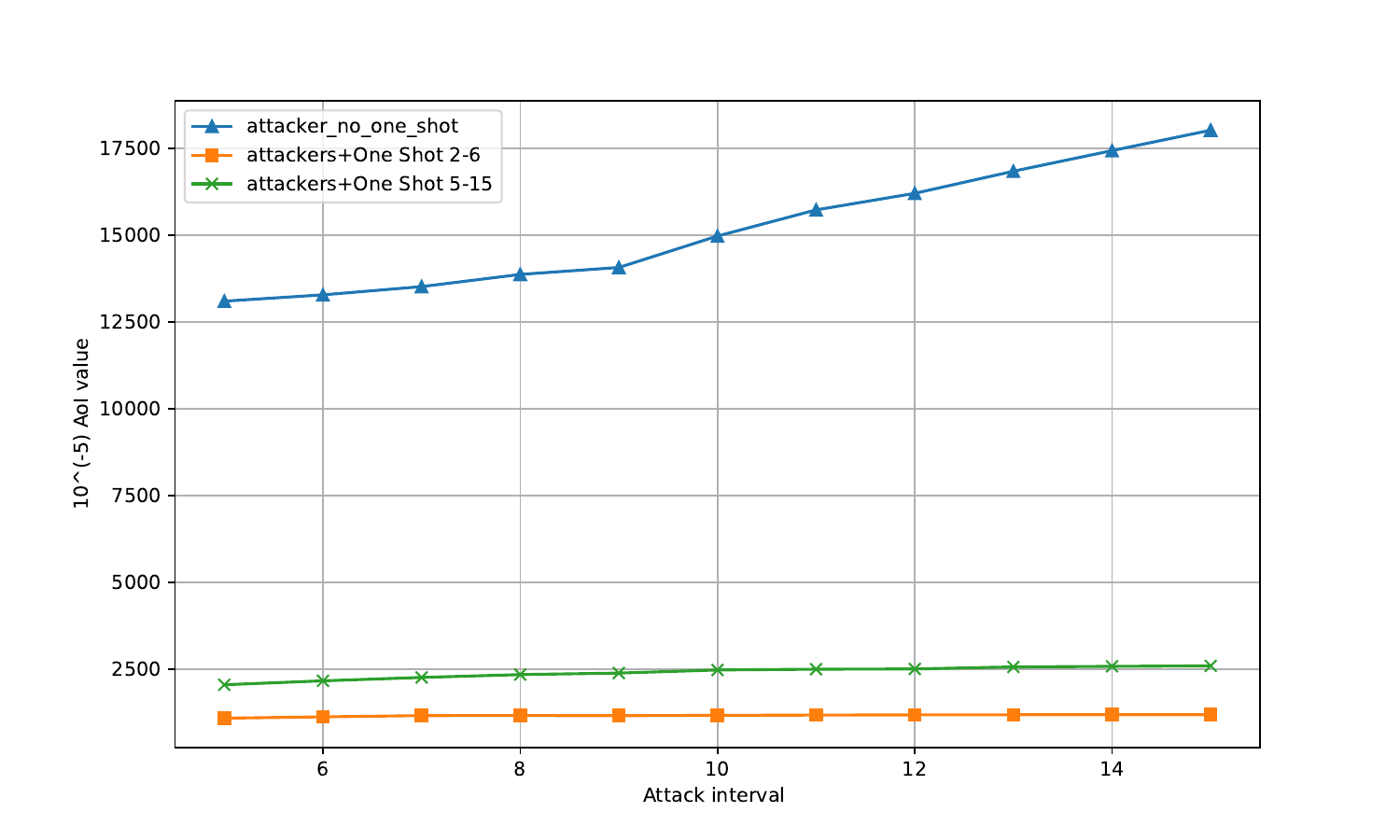} \\
		\label{ann_images}
      \caption{Impact of Attacker Interval on $10^{-5}$ AoI tail, 5 targets}

	\end{figure}

 \begin{figure}[!h]
		\centering
		\includegraphics[width=0.35\textheight]{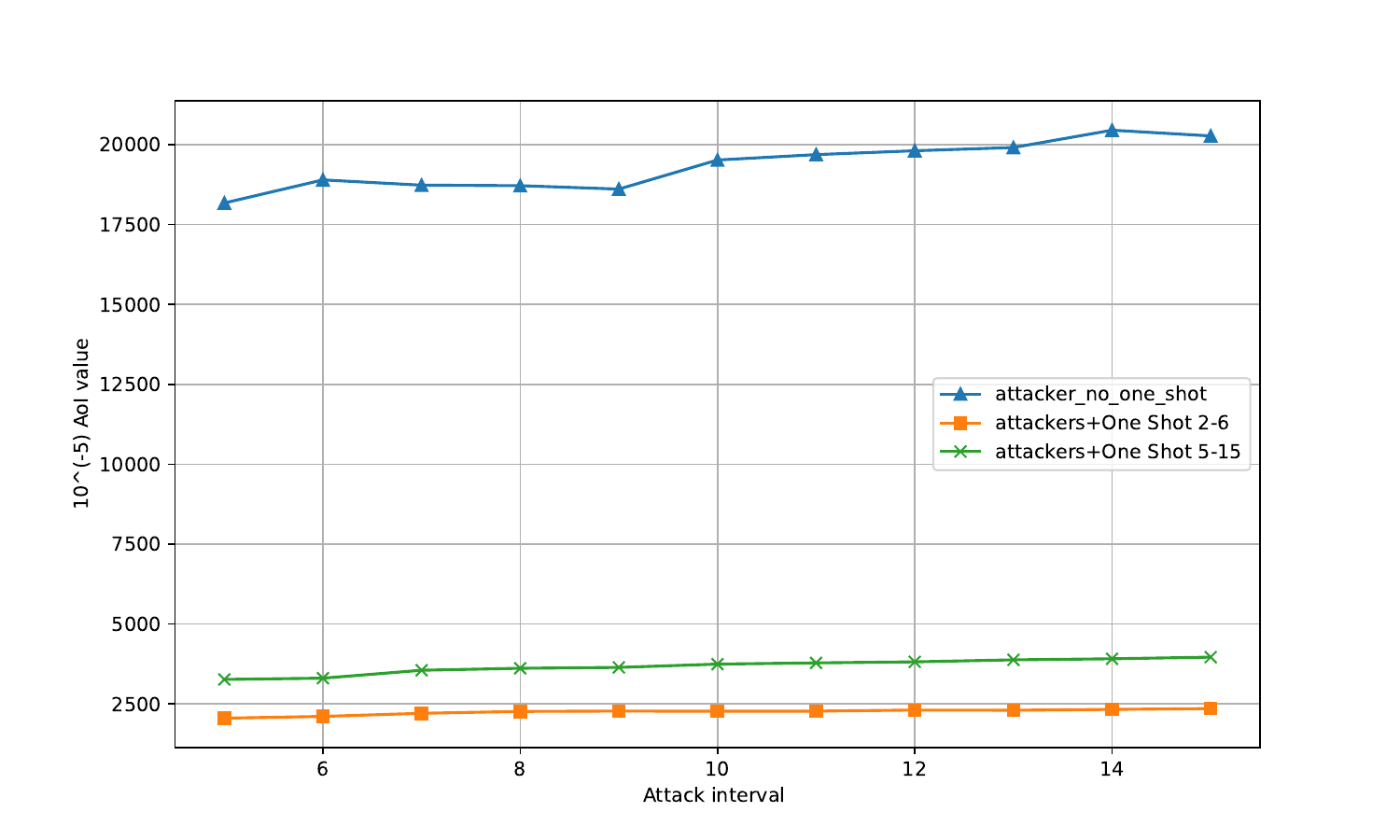} \\
		\label{ann_images}
      \caption{Impact of Attacker Interval on $10^{-5}$ AoI tail, 30 targets}

	\end{figure}
\section{Conclusion}
This study demonstrates that the one-shot Semi-Persistent Scheduling (SPS) mechanism significantly enhances the resilience of Cellular Vehicle-to-Everything (C-V2X) networks against smart Denial-of-Service (DoS) attacks. Through Monte Carlo simulations, the one-shot approach is shown to improve Packet Delivery Ratio (PDR), reduce Inter-Packet Gap (IPG), and lower Age of Information (AoI), particularly in low-density vehicular environments. While the benefits decrease slightly in high-density scenarios, the mechanism remains a robust solution for maintaining communication stability under attack. The findings validate the one-shot SPS as an effective strategy for improving the security and reliability of C-V2X networks.


%




\ifCLASSOPTIONcaptionsoff
  \newpage
\fi

\end{document}